\documentclass[conference]{IEEEtran}
\ifCLASSINFOpdf
\else
\fi
\usepackage{verbatim}
\usepackage{graphicx}
\usepackage{epstopdf}
\usepackage{multirow}
\usepackage{textcomp}
\usepackage{adjustbox}
\begin{document}
%
\title{Deep Neural Network Based Precursor microRNA Prediction on Eleven Species}

\author{Jaya Thomas $^{a,b}$, Lee Sael$^{a,b}$*}
\author{\IEEEauthorblockN{Jaya Thomas}
\IEEEauthorblockA{ $^{1}$Department of Computer Science, \\ Stony Brook University, \\Stony Brook, NY 11794, USA\\
$^{2}$Department of Computer Science, \\ State University of New York Korea, \\ Incheon 406-840, Korea;\\
Email: jaya.thomas@sunykorea.ac.kr}

\and
\IEEEauthorblockN{Lee Sael}
\IEEEauthorblockA{$^{1}$Department of Computer Science, \\ Stony Brook University, \\Stony Brook, NY 11794, USA\\
$^{2}$Department of Computer Science, \\ State University of New York Korea, \\ Incheon 406-840, Korea;\\
Email: sael@cs.stonybrook.edu}

}

%


\maketitle

\begin{abstract}
MicroRNA (miRNA) are small non-coding RNAs that regulates the gene expression at the post-transcriptional level. Determining whether a sequence segment is miRNA is experimentally challenging. Also, experimental results are sensitive to the experimental environment. These limitations inspire the development of computational methods for predicting the miRNAs.
We propose a deep learning based classification model, called DP-miRNA, for predicting precursor miRNA sequence that contains the miRNA sequence.
The feature set based Restricted Boltzmann Machine method, which we call DP-miRNA, uses 58 features that are categorized to four groups: sequence features, folding measures, stem-loop features and statistical feature.
We evaluate the performance of the DP-miRNA on eleven twelve data sets of varying species, including the human.
The deep neural network based classification outperformed support vector machine, neural network, naive Baye's classifiers, k-nearest neighbors, random forests, and a hybrid system combining support vector machine and genetic algorithm.
\end{abstract}


%

\section{Introduction}
\par MicroRNAs (miRNAs) are single-stranded small non-coding RNA that are typically $sim$22 nucleotides long. The main functionality of the miRNA is the  translation regulation of the messenger RNAs (mRNA). The miRNA regulates gene expression at the post transcription level by base pairing with the complementary sequence. This process hinders the translation of mRNA to proteins. The regulatory role of miRNAs are important in development, cell proliferation and cell death.  Thus, their deregulation has been connected with neuro-degenerative disease, cancer and metabolic disorders \cite{Witkos2011}. Moreover, miRNA can bind to multiple target mRNAs, making the determination of true miRNA even more crucial.

\par The miRNA biogenesis involves number of steps. First, primary transcripts of miRNA (pri-miRNA) are transcribed from introns of protein coding genes that are several kilobases long. The pri-miRNAs are then clopped by Rnase-III enzyme Drosha into $\sim$70 base pairs (bp) long hair-pin-looped precursor miRNAs (pre-miRNAs).
Exportin-5 proteins transport pre-miRNA hairpins into the cytoplasm through nuclear pore. In cytoplasm, pre-miRNAs are further cleaved by Rnase-III enzyme Dicer to produce a $\sim$20 bp double stranded intermediate called miRNA:miRNA*.
A strand of the duplex with the low thermodynamic energy becomes a mature miRNA.
Most mature miRNAs interact with the RNA interference (RNAi) induced silencing complex (RISC) through base pairing of the target mRNAs and regulate the expression of the genes.

\par Currently,  miRBase \cite{Zhong2015} reports over 28645 miRNAs in more than 200 species, out of which over 2000 miRNA are reported for human. Informatics analysis predicts that 30$\%$ of human genes are regulated by miRNA \cite{Ross2007}. miRNAs can be experimentally determined by directional cloning of endogenous small RNAs \cite{Chen2005}. However, this is a time consuming process that require expensive laboratory reagents. These drawbacks motivate the application of computational approaches for predicting miRNAs.

\par Machine learning based methods can identify nonhomologous and species-specific miRNAs as compared to homologous search and comparative genomics approaches \cite{Zhong2015}. Distinguishing real pre-miRNAs and other pseudo hairpins is a problem that can be readily expressed as a binary classification problem.
In this context, the human pre-miRNAs are labeled as ‘+1’ or positive samples whereas pseudo hairpins are labeled as ‘-1’ or negative samples. The derived features are learned and mapped to the feature space for classification. Many approaches have been developed using artificial neural networks (ANN)\cite{Rahman2012}, support vector machines (SVM), and random forests (RF)\cite{Xiao2011}.
Among them, SVM has been most extensively applied. Some of the notable SVM-based methods including triplet-SVM\cite{Xue2005}, MiRFinder \cite{Huang2007}, miPred \cite{Ng2007}, microPred \cite{Batuwita2009}, yasMiR \cite{Pasaila2011}, YamiPred \cite{Kleftogiannis2015}, MiRenSVM \cite{Ding2010}, and MiRPara \cite{Wu2011}. However, no Deep neural network based methods were explored.

\par Deep neural network (DNN) algorithm performs well in a setting where there are no obvious representative features that can be extracted from raw data. Convolution neural network (CNN) has been used in several instances to directly process raw data as input. deep restricted boltzmann machine (RBM), on the other hand has been popular with where there are large number of features. Whether the input is a raw data or a high dimensional feature, DNN uses multi-layer architecture to infer from data. The deep architecture automatically extracts high-level feature necessary for classification or clustering. The multiple layers in deep learning helps exploit the inherent complexities of data. DNN have exhibited a good performance in different machine learning problems such as protein structure prediction {\cite{Spencer2015}}, and  predict splicing patterns {\cite{Leung2014}}.

\par In this work, we utilize heterogeneous features including sequence features, folding measures, stem-loop features and statistical features (z-score) to differentiate pre-miRNAs from pseudo hairpins.
The pseudo hairpins are RNA sequences, which have similar stem-loop features to pre-miRNAs but does not contain mature miRNAs.
We use experimentally validated pre-miRNAs as positive examples and pseudo hairpins as negative examples to train and test the proposed method.
We compared the performance of proposed model against existing machine learning classifier on eleven different species which extends the previous work on human dataset only \cite{Thomas2017}. 

The main contribution of the paper are summarized as follows:
\begin{itemize}
\item{Deep learning based prediction model is proposed for integrating large number of heterogeneous features for predictive analysis of pre-miRNAs from pseudo hairpins.}
\item{Modified sampling technique is applied to address class imbalance problem.}
\item{We apply the method on eleven different species.}
\end{itemize}

\section{Background}

\par Machine learning approaches are the most commonly used for miRNA prediction. Many tools have been developed based on the different classification techniques such as naive Bayes classifier (NBC), artificial neural networks (ANN), support vector machines (SVM), and random forests (RF). The most widely used technique amongst them is SVM, several tools are build using it, which includes triplet-SVM\cite{Xue2005}, MiRFinder \cite{Huang2007}, miPred \cite{Ng2007}, microPred \cite{Batuwita2009}, yasMiR \cite{Pasaila2011}, YamiPred \cite{Kleftogiannis2015}, MiRenSVM \cite{Ding2010} and MiRPara \cite{Wu2011}.

\par Amongst SVM the most commonly used classifier is G$^{2}$DE \cite{Hsieh2010}, a kernel based machine learning approach with higher prediction accuracy than existing kernel and logic based classifiers. For triplet-SVM\cite{Xue2005}, the classifier distinguish the hairpins of real pre-miRNAs and pseudo pre-miRNAs. They consider local structure-sequence features that reflect characteristics of miRNAs. The approach report an accuracy of 90$\%$ considering pre-miRNAs from the other 11 species including plants and virus without considering any other comparative genomics information. Another, miPred\cite{Ng2007} SVM approach considered Gaussian Radial Basis Function kernel (RBF) as a similarity measure for  global and intrinsic hairpin folding attribute and resulted with accuracy of around 93$\%$. MicroPred\cite{Batuwita2009} introduce some additional features for evaluation of miRNA using SVM based machine learning classifier. Author's report classification results of high specificity of 97.28$\%$ and sensitivity of 90.02$\%$. The miR-SF classifier \cite{Wang2011}, predict the identified human pre-miRNAs in miRBase source on the selected optimized feature subset including 13 features, generated by SVM and genetic algorithm. YamiPred \cite{Kleftogiannis2015}, is a genetic algorithm and SVM based embedded classifier that consider feature selection and parameters optimization for improving performance.

\par  Some studies have used ANN machine learning technique, as in MiRANN\cite{Rahman2012} predictor, author's consider ANN for pre-miRNA prediction by expanding the network with more neurons and the hidden layers.  The network is designed to be impartial for any feature by integrating exceptional weight initializing equation where closest neurons slightly differ in weights.

\section{Methods}

\subsection{Data}
\par The human pre-miRNA sequence was retrieved from the miRBase 18.0 release.
Similar to miPred \cite{Ng2007} approach, the multiple loops were discarded to get 1600 pre-miRNA as positive dataset.
The obtained sequence had an average length of 84 nt with minimum 43 nt and maximum 154 nt. The negative dataset consists of 8494 pseudo hairpins as the false samples.
They were extracted from the human protein-coding regions as suggested by microPred \cite{Batuwita2009}.
The average length of the sequence is 85 nt with minimum as 63 nt and maximum as 120 nt.
The different filtering criteria, including non-overlapping sliding window, no multiple loops, lowest base pair number set to 18, and minimum free energy less than 15kcal/mol were applied on these sequences to resemble the real pre-miRNA properties.

\par To evaluate the performance of the proposed model, we also consider miRNA data for 10 other species obtained from Kleftogiannies at el. \cite{Kleftogiannis2015}. Some of these species are similar to human including gorillas and chimpanzees, where as others are distinct including Aves and Rodentia. The number of positive and negative data samples for considered species are summarized in Table ~\ref{table:00}. A more detailed description on the dataset can be found in the original article \cite{Kleftogiannis2015}.

\par The training set (TE-H) were randomly selected 200 human pre-miRNAs from miRBase 8.2, and 400 pseudo-miRNA considered in TripleSVM.  Here, TE-H dataset contains 123 human  pre-miRNAs remaining from miRBase 8.2 after allocation of pre-miRNA for training (TR-H).
The IE-NH dataset included 1918 pre-miRNA from 40 non-human species from miRBAse 8.2, IE-NC consist of 12387 non-coding RNAs from Rfam 7.0 database \cite{Griffiths-Jones2005}, and IE-M included 31 messenger RNAs selected from GenBank \cite{Benson2007}.

\begin{table*}[t]
\caption{Class description for different species}
\label{table:00}
\centering
\hspace*{-3em}
 \begin{tabular}{llllllllllll}
    \hline
      \multicolumn{11}{c}{Species} & \\
            Class &Human&Caballus&Canis&Gallus&Gorilla&Mus&Panciscus&Pongo&Taurus&Troglodytes&Virus\\
  \hline
Positive&	1600&341&323&497&85&720&88&581&662&599&237 \\
Negative&	8494&517&520&2000&1000&3000&116&1356&3000&176&107\\

\hline
\end{tabular}
\end{table*}

\subsection{Feature set}

\par The common characteristics of pre-miRNAs used for evaluation consists of  sequences composition properties, secondary structures, folding measures and energy.  This work adopts 58 characteristic features, which are shown useful in existing studies for predicting miRNA. The sequence characteristics include features related to the frequency of two and three adjacent nucleotide and aggregate dinucleotide frequency in the sequence. The secondary structure features from the perspectives of miRNA bio-genesis relating different thermodynamic stability profiles of pre-miRNAs. These structures have lower free energy and often contain stem and loop regions. They include diversity, frequency, entropy-related properties, enthalpy-related properties of the structure. The other features are hairpin length, loop length, consecutive base-pairs and ratio of loop length to hairpin length of pre-miRNA secondary structure. The energy characteristic associated to the energy of secondary structure includes the minimal free energy of the secondary structure, overall free energy NEFE, combined energy features and the energy required for dissolving the secondary structure.

The description of these features is summarized in table ~\ref{table:00}.

\begin{table*}

\caption{Features for miRNA Prediction}
\centering
\label{table:00}

\begin{adjustbox}{max width=\textwidth}
 \begin{tabular}{lll}
    \hline

             Feature&Number&Description\\
  \hline
XY, where X,Y$\in$ $\{$A,C,G,U$\}$&16&	Dinucleotide pairs frequency\\
XYZ, where X,Y,Z$\in$ $\{$A,C,G,U$\}$&64&	Trinucleotide pairs frequency\\
A+U$\%$&1&Aggregate dinucleotide frequency (bases which are either A or U)\\
G+C$\%$&1&Aggregate dinucleotide frequency (bases which are either G or C)\\
L&1& Structure length\\
Freq&1&Structural frequency property \\
dP&1&Adjusted base pairing propensity given as  total$\_$bases$/$L\\
dG&1&Adjusted Minimum Free Energy of folding given as dG = MFE$/$L\\
dD&1&Adjusted base pair distance \\
dQ&1& Adjusted shannon entropy \\
dF&1&Compactness of the tree-graph representation of the sequence\\
MFEI1&1&MFEI1 = dG$/$$\%$(C+G)\\
MFEI2&1&MFEI2 = dG$/$number$\_$of$\_$stems\\
MFEI3&1&MFEI3 = dG$/$number$\_$of$\_$loops\\
MFEI4&1&MFEI4 = dG$/$total$\_$bases\\
MFEI5&1& MFEI5= dG$/$$\%$(A+U) ratio\\
 dS&1&Structure entropy\\
 dS$/$L&1&Normalized structure entropy\\
dH&1&Structure Enthalpy \\
dH$/$L&1&Normalized structure enthalpy \\
Tm&1& Melting Temperature \\
Tm$/$L&1&Normalized melting temperature \\
BP$/$X, where X $\in$ $\{$GC,GU,AU$\}$&3& Ratio of total$\_$bases to respective base pairs\\

 G$/$C &1&Number of G,C bases\\
Avg$\_$BP$\_$Stem&1&Average number of base pairs in the stem region \\
$\mid$A$-$U$\mid$$/$L,$\mid$G$-$C$\mid$$/$L, $\mid$G$-$U$\mid$$/$L&3& $\mid$X$-$Y$\mid$ is the number of (X $-$Y) base pairs in the secondary structure\\
$\mid$A$-$U$\mid$$\%$$/$n$\_$stems, $\mid$G $-$ C$\mid$$\%$$/$n$\_$stems, \\and $\mid$G $-$ U$\mid$$\%$$/$n$\_$stems
&3&Average number of base pairs in the stem region\\
zP,zG,zD,zQ,zSP&5&statistical Z-score of the folding measures\\
dPs &1& Positional Entropy which estimates the structural volatility of the secondary structure\\
EAFE&1&Normalized Ensemble Free Energy\\
CE$/$L&1& Centroid energy normalized by length\\
Diff&1& Diff =$\mid$ MFE-EFE$\mid$$/$L where, EFE is the ensemble free energy\\
IH&1&Hairpin length dangling ends\\
IL&1& Loop length\\
IC&1& Maximum consecutive base-pairs\\
$\%$L&1&Ratio of loop length to hairpin length\\
\hline
\end{tabular}
\end{adjustbox}
\end{table*}

\begin{table*}[t]

\caption{Comparison with existing computational intelligence techniques} 
\centering 
\begin{tabular}{p{2cm} p{2cm} p{2cm} p{2cm} p{2cm}} 
\hline\hline 
Classification Method & Accuracy & Sensitivity & Specificity & Geometric Mean \\ [0.5ex] 
\hline 

NBC&0.914 $\pm$0.003	&0.943 $\pm$0.003&	0.796 $\pm$0.012&	0.867$\pm$ 0.006\\
KNN&	0.908$\pm$ 0.005&	0.970$\pm$ 0.122&	0.657$ \pm$0.023&	0.798$\pm$ 0.009\\
RF&	0.937$ \pm$ 0.004&	0.979$ \pm$0.002&	0.765$ \pm $0.002&	0.865$\pm $0.008\\
miRANN&0.917$ \pm$ 0.002&	0.963$ \pm$0.004&	0.705$ \pm $0.006&	0.837$\pm $0.006\\
YamiPred&	0.932$ \pm $0.005&	0.937$\pm $0.008&	0.912$ \pm $0.012&	0.924$ \pm$0.004\\
DP-miRNA&	0.968$ \pm $ 0.002	&0.973$ \pm $ 0.005&	0.942$ \pm $ 0.006&	0.971$ \pm $ 0.004\\
\hline 
\end{tabular}
\label{table:01} 
\end{table*}

\subsection{Deep neural network}
{
\begin{figure}[!ht]
  \centering
  \includegraphics[clip,trim=0cm 0.1cm 0cm 0cm,width=.95\columnwidth]{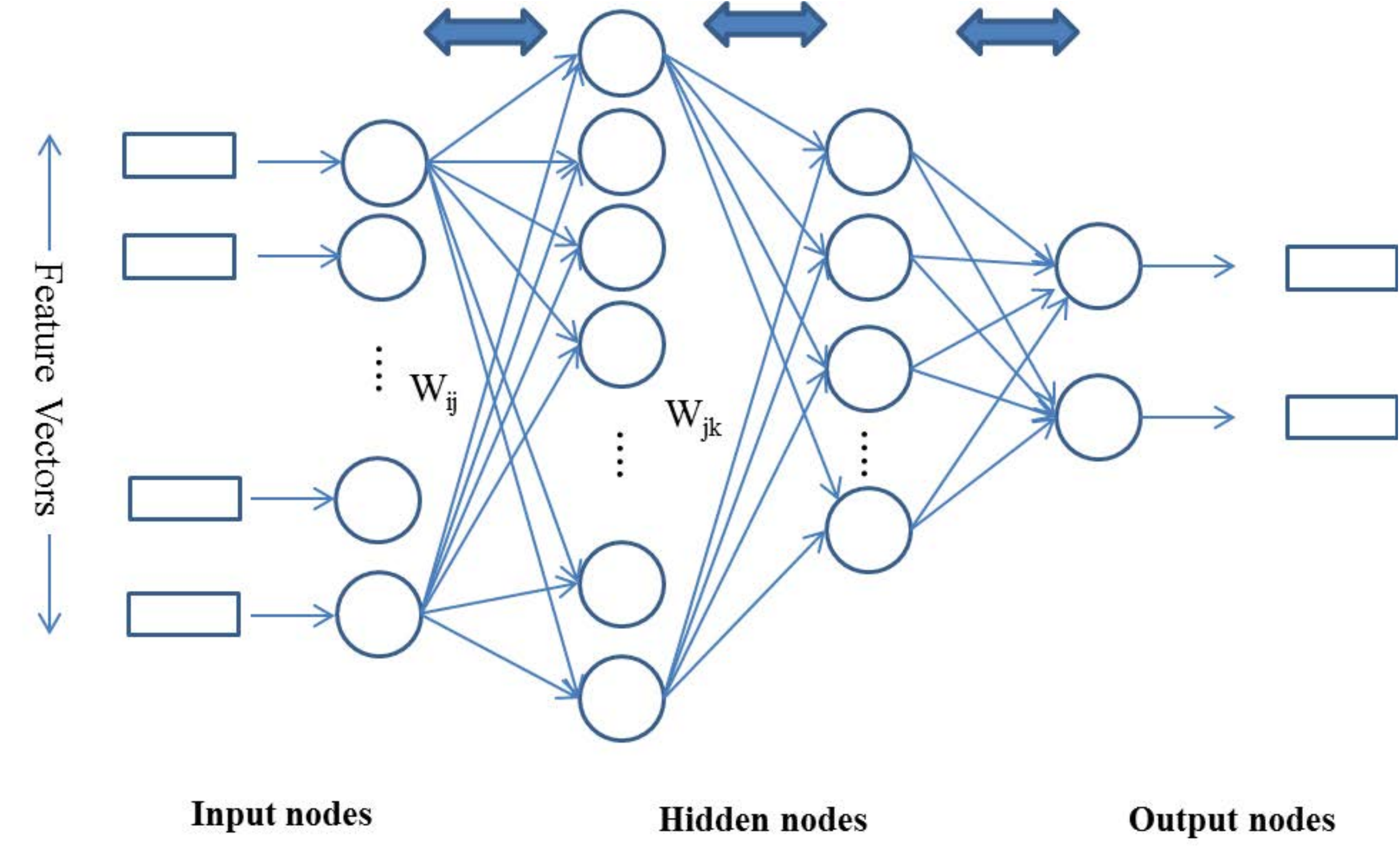}
  \caption{A deep learning to predict miRNA with exracted features}
\label{fig:01}
\end{figure}
}

\par The proposed deep neural network (DNN) based miRNA prediction method, we call DP-miRNA, has three hidden layers, and the model is denoted as  X-100-70-35-1, where X being the size of the input layer, 1 denotes the number of neuron in the output layer and the remaining values denotes the number of neurons in each hidden layer. Figure 1 illustrates the model architecture and layer-by-layer learning procedure. Different model architectures were trained using the same learning procedure but varying the number of hidden layer and nodes. Amongst the candidate network models, a better one was selected based on the classifier accuracy. The network model is pre-trained layer after the layer with the restricted Boltzmann machine (RBM). The initialization of the weight between every pair of adjacent layer is a step process that begins from the input or visible layer and completes at last hidden layer. At first, RBM learns the structure of the input data that constitutes to the activation of the first hidden layer, then the data is moved one layer down the network. Going in reverse, with each new hidden layer, the input from the previous layer is approximated by adjusting the network weights. This back and forth adjustment process is termed as Gibbs sampling, where the weights are updated by considering the difference in the correlation of the hidden activations and visible inputs. The process continues and now the first hidden layer will act as the input, which is multiplied by weights at the nodes of second hidden layer and the probability for activating the second hidden layers is calculated. This process results in sequential sets of activations by grouping features of features resulting in a feature hierarchy, by which networks learn more complex and abstract representations of data. This procedure of training a RBM can be repeated several times to create a multi-layer network.  At the end, a standard feed-forward neural network is added after the last hidden layer, so the input being the activation probabilities which is used to predict the label. The resulting deep network was put together to adjust the weights using the standard back propagation algorithm to minimize the cross-entropy cost function error \cite{Hinton2002}.


The deep learning network are trained with standard back propagation algorithm, with the weights adjusted using the stochastic gradient descent as \cite{Hinton2012}]:
\begin{equation}
w_{ij}(t+1)=w_{ij}(t) + \eta \frac{\partial C}{\partial w_{ij}},
\end{equation}
where, $ w_{ij} (t+1)$ is the weight computed at $t+1$,  $\partial$ denotes the learning rate, and C is the cost function.  For the given model, softmax is used as an activation function and the cost is computed using cross entropy. The softmax function is defined as
\begin{equation}
p_{j}=\frac{exp(x_{j})}{ \sum_{k}exp(x_{k})}.
\end{equation}
Here, $p_{j}$ stands for the output of the unit j,  $x_{j}$ and $x_{k}$ denotes the total input to unit j and k respectively for the same level. The cross entropy is given by
           \begin{equation}
C=-\sum_{i} d_{j} log(p_{j}),
\end{equation}
where $d_j$ is the target probability for output unit j and $p_j$ is the probability output after applying the activation function.

\subsection{Class imbalance}
Another problem that we have addressed here is the class imbalance problem in miRNA predictions. Class imbalance is a machine learning problem where the number of data samples belonging to one class (positive) is far less compared to data sample belonging to other class (negative). The imbalance class is often solved by using either under or over sampling methods. In case of under sampling the data samples are removed from the majority class, whereas for over sampling balance is created either by addition of new samples or duplication of the existing minority class samples.
Class imbalance problem arise in miRNA data classification because in cells the quantity of molecules, which are not pre-miRNAs fold into a miRNA-like shape that is larger than the real miRNA. In existing classifiers such as triplet-SVM\cite{Xue2005} and  miPred \cite{Ng2007} handled the imbalance problem manually.

We address the class imbalance problem during the training phase by adopting a modified under sampling approach \cite{Yen2009}. In the modified approach, we divided the majority class into subsets using k-means algorithm with k=5, and thus obtain clusters with slightly higher similarity amongst the group. These clusters are used to form different training sets by varying the ratio of majority class sample to minority class samples. Amongst the training data set, one with higher accuracy was selected as an input to the classifier.

\section{Results}

To evaluate the performance of the proposed classifier, we compare our method to existing state of the art miRNA classifiers. The evaluation is carried out by dividing the available data samples into training (60\%), validation (20\%) and testing (20\%) set. The size of the input vector here is 58, i.e., the number of features used to build the model. The input data is normalized to standardizing the inputs in order to improve the training and to avoid getting stuck in local optima.

\subsection{ Performance evaluation metrics}
\par The DP-miRNA model is a two class classifier, where true positive (TP) denotes the number of data samples classified as positive (real pre-miRNAs) and true negative (TN) represent  correctly classified negative samples (pseudo pre-miRNAs). Similarly false positive(FP) and false negative (FN) represents the numbers of the misclassified positive and negative samples, respectively. The other measuring terms are sensitivity (SE) that measures the proportion of positives that are correctly identified accounting for the total positive samples, SE=TP/(TP+FN). Specificity (SP) measures the proportion of negatives that are correctly identified accounting for the total negative sample, SP =TN/(TN + FP). The classification accuracy (Accuracy) is proportion of correctly classified positive and negative class samples to total number of samples, Accuracy= (TP + TN)/(TP + TN + FP + FN). Another measure is geometric mean (Gm) to evaluate global classification performance, Gm=$\sqrt{SE \times SP}$.

\subsection{DP-miRNA model selection}
\par The DP-miRNA classifier learns  more abstract features from the lower one to better summarize the pre-miRNAs and pseudo hairpins in the vector space.
Table ~\ref{table:01} shows a comparative result of the proposed DP-miRNA against the common machine learning approach for miRNA prediction. Considering the stochastic nature of the algorithm the output values are averaged for twenty executions. In comparison to the tested machine leaning techniques, DP-miRNA classifier shows a better performance. Another, point observed was that the modified sampling approach helped to overcome class imbalance problem as compared to random selection of data during training phase.

Figure ~\ref{fig:01}  illustrates the deep learning network for layer-by-layer learning process. We considered the model architecture for deep learning as X-100-70-1, where X is the size of the feature vector. The model architecture was selected after experimental validation. Initially, different model architectures were trained using the same learning procedure, but by varying the parameters including number of hidden layer, the number of hidden nodes and the epoch for each training process. The different model obtained were evaluated for the training error. It was observed that after insertion of two hidden layers, an addition of new hidden layers resulted in no significant improvement in error rate.

We compared the performance difference between the two models when the number of hidden nodes and layers are varied for gorilla species data set.
The performance for the two models shows that by increasing the number of hidden neuron both the training error and the execution time is increased.
These experimental evaluation helped us to select an appropriate model architecture during training. Based on the performance, we modeled the the DP-miRNA with two hidden layer.


\par Further, we examined the performance of the
DP-miRNA on selected twenty features that mostly represented sequence information and other thermodynamical characteristics. The feature set consist of dinucleotide frequencies AG, AU, CU, GA, UU, MFEI4, MFEI5, Positional Entropy, EAFE, Freq, dH/L, Tm, Tm/L, L, Avg$\_$BP$\_$stems, (G-U)/stems, (CE/L), (A-U)/stems and Statistical Z-scores zG, zQ and zSP. On the selected feature set we obtained a accuracy of 99.2$\%$, with sensitivity and specificity high as 99.58$\%$ and 98.24$\%$ respectively. The result support the fact that features as entropy, enthalpy, minimum free energy and melting temperature are crucial for predicting miRNA \cite{Kleftogiannis2015}.

\subsection{Predictability evaluation using 10 species}

\begin{figure}[!ht]
  \centering
 \includegraphics[clip,trim=0cm 0.1cm 0cm 0cm,width=0.5\textwidth]{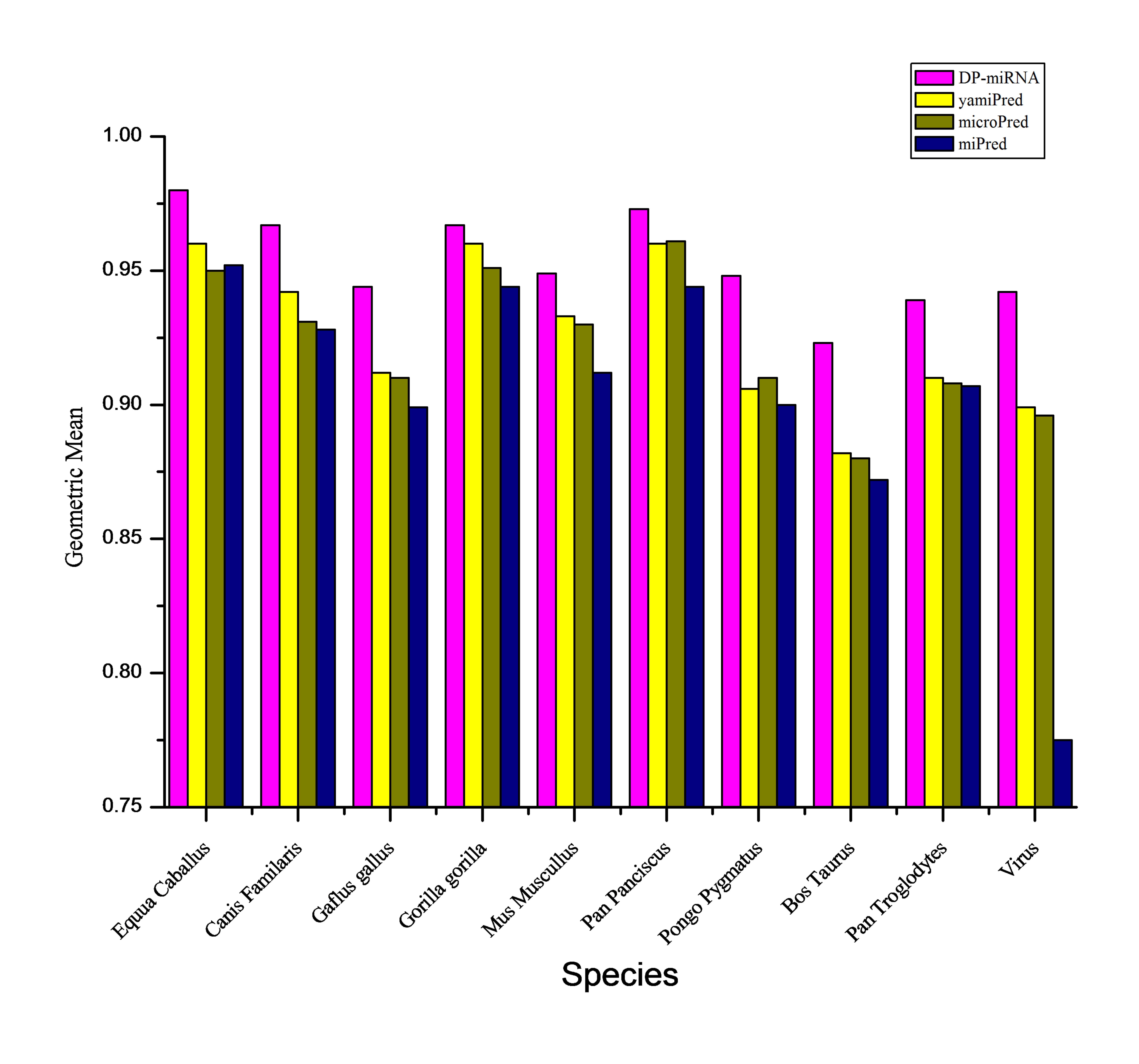}
  \caption{Performance evaluation for different species}
\label{fig:03}
\end{figure}

\par We evaluate the performance of the proposed classifier for miRNA data of the different species. Further, we compare the obtained results with the existing classifier as YamiPred, miPred and microPred. We carry out the comparison for ten different organism. The classification results obtained can be refered in Figure ~\ref{fig:03}.
We observed that the model build using deep learning was able to achieve a better generalization ability.
The performance accuracy for Bos Taurus species was observed less as compared to other species.
The model was initially evaluated with all selected features and obtained geometric mean of  89.62$\%$.
We modified the feature set to include the features specific to Bos Taurus data. The new feature set included 11 ($\%$AþU, AU, CU, UU, MFEI4, EAFE, dH/L, Tm/L, j G-C j /L, (G-C)/stems, ZG), 4 (AU, EAFE, dH/L, Tm/L), and additional characteristics such as more dinucleotide frequencies (AC, GA, GG, CA, UG), statistical score ZP, topological factor dF and BP/AU, BP/GU. After the feature modification, mean performance improved from  previous 89.62$\%$ percent to 92.3$\%$.

\section{Conclusion}
\par In this paper, we proposed deep learning classifier based pre-miRNA prediction method and showed performance improvement over existing methods. The proposed classifier was evaluated extensively on human dataset and ten other species. The 58 features used as the input to deep learning framework included sequence conservation features, secondary structure features, and energy features of miRNA. For comparison, the dataset were generated with four biologically significant groupings and their combined set.

\section*{Acknowledgement}
This research was supported by Basic Science Research Program through the RNF of Korea (NRF-2015R1C1A2A01055739), by the KEIT Korea under the "Global Advanced Technology Center" (10053204) and by the MSIP, Korea, under the "ICT Consilience Creative Program" (IITP-2015-R0346-15-1007) supervised by the IITP.


\bibliography{mybibfile}

\end{document}